\documentclass[twocolumn,showpacs,aps,prl,superscriptaddress,
groupedaddress]{revtex4} 
\usepackage{graphicx}  
\usepackage{dcolumn}   
\usepackage{bm}        
\usepackage{amssymb}   
\usepackage{subfigure} 
\graphicspath{{figures/}}
\begin{document}

\title{Hydrodynamic Correlations slow down Crystallization of Soft Colloids}

\author{D.~Roehm} 
\email{dominic.roehm@icp.uni-stuttgart.de}
\affiliation{University of Stuttgart,
Allmandring 3, 70569 Stuttgart, Germany}

\author{S.~Kesselheim} 
\email{stefan.kesselheim@icp.uni-stuttgart.de}
\affiliation{University of Stuttgart,
Allmandring 3, 70569 Stuttgart, Germany}

\author{A.~Arnold}
\email{axel.arnold@icp.uni-stuttgart.de}
\affiliation{University of Stuttgart, Allmandring 3,
70569 Stuttgart, Germany}

\date{\today}

\begin{abstract}
  Crystallization is often assumed to be a quasi-static process that is
  unaffected by details of particle transport other than the bulk diffusion
  coefficient. Therefore colloidal suspensions are frequently argued to be an
  ideal toy model for experimentally more difficult systems such
  as metal melts. In this letter, we want to challenge this assumption. To
  this aim, we have considered molecular dynamics simulations of the 
  crystallization in a suspension of Yukawa-type colloids. In order to
  investigate the role of hydrodynamic interactions (HIs) mediated by the
  solvent, we modeled the solvent both implicitly and explicitly, using Langevin
  dynamics and the fluctuating Lattice Boltzmann method, respectively.  Our
  simulations show a dramatic reduction of the  crystal growth velocity due to
  HIs even at moderate hydrodynamic coupling. A detailed analysis shows that
  this slowdown is due to the wall-like properties of the crystal surface,
  which reduces the colloidal diffusion towards the crystal surface by
  hydrodynamic screening. Crystallization in suspensions therefore
  differs strongly from pure melts, making them less useful as a toy
  model than previously thought.
\end{abstract}

\maketitle

Crystallization and nucleation of undercooled melts are often studied
using model systems of charged colloids in solution~\cite{hamaguchi96a,
lowen96a, loewen97a}, such as polystyrene (PS) or polymethylmethacrylat (PMMA)
spheres suspended in water~\cite{klein89a, petsev92a, engelbrecht11a}. The
solvent gives rise to hydrodynamic interactions (HIs) between the colloidal
particles that are not present in, e.g., metal melts. The influence of HIs on
the dynamical properties of colloidal suspensions has been extensively
studied in recent years~\cite{naegele96a, padding06a}. L\"owen
\textit{et al.}~\cite{lowen93a,lowen96a} showed that the ratio of the
long-time to short-time self-diffusion coefficients has a universal
value along the fluid freezing line. Recent studies by Pesche~\cite{pesche01a}
and N\"agele~\cite{mcphie07a} of quasi-2D dispersions
show that HIs have an impact on the self-diffusion function in these
soft-sphere suspensions. However, since crystal growth happens on much
larger time scales, it is commonly believed to be a quasi-static
process that is unaffected by HIs. For example, in classical nucleation
theory (CNT)~\cite{kalikmanov13a}, hydrodynamic interactions are assumed to
enter only through the effective diffusion constant of the attaching
colloids, which can be measured conveniently in the bulk liquid. Also
in most computer simulation studies of nucleation, hydrodynamic
interactions are neglected to avoid the high computational costs.

In the following, we will show with the help of computer simulations
that HIs do have a remarkable influence on the dynamics of the
crystallization in a colloidal suspension. Even at moderate coupling, 
we found the crystal growth velocity to be reduced by a factor of three.
Similar findings have been reported by Schilling \emph{et al.} using different
simulations techniques~\cite{radu13a}.


To investigate the influence of hydrodynamic interactions on crystal
growth, we studied the crystallization of Yukawa-type particles
confined between two planar walls. The influence of hydrodynamic
correlations on crystallization was evaluated by performing
both simulations including and excluding HIs, by employing a fluctuating
lattice Boltzmann method~\cite{duenweg09a} and Langevin
dynamics~\cite{hinch75a}, respectively. We prepare our systems as an
undercooled liquid and let the system crystallize. Due to the presence of the
confining walls, the nucleation barrier is sufficiently small, so that we do
not need special rare event sampling techniques. All simulations were performed
using the MD simulation package
\emph{ESPResSo}~\cite{arnold04b,roehm12a,arnold13a}.

As interparticle potential we used a screened Coulomb interaction potential
\begin{equation}
 U(r)=l_Bk_BT\frac{Q_{1} Q_{2} \exp(-\lambda_D
 r)}{r},
  \label{equ3}
\end{equation}
where $l_B$ is the Bjerrum length, $k_B$ is the Boltzmann constant, $Q_1$ and
$Q_2$ give the charges on the interacting particles,
$r$ is distance between the particles. The range of the potential is
determined by the Debye-H\"uckel screening
length $\lambda_D$. The static properties of a Yukawa system can be
characterized by two dimensionless parameters~\cite{hamaguchi96a}
\begin{equation}
  \kappa = \frac{w}{\lambda_D} \quad\text{and}\quad
  \Gamma = \frac{Q_1 Q_2}{4\pi \epsilon_0 w k_b T}
  = \frac{Q_1 Q_2 l_B}{w},
\end{equation}
where $w=(3/(4\pi \rho))^{1/3}$ is the Wigner-Seitz radius of the crystal
phase and $\rho$ the
particle density. Phase diagrams of systems with Yukawa-type
interactions have been calculated both by Monte Carlo
simulations~\cite{meijer91a} and MD
simulations~\cite{hamaguchi96a,hamaguchi97a},
which consistently found three regimes: a fluid phase and two
different solid phases with BCC or FCC structure, respectively. For our
simulations, we chose $\kappa=3.0$ and $\Gamma=$1260, which is
slightly above the fluid-solid transition line in the BCC
regime.
The walls act on the particles via a Weeks-Chandler-Andersen (WCA) potential.
When modeling the solvent by a Langevin thermostat~\cite{andersen80a},
drag and random forces on the particles lead to the correct thermal
distribution, but hydrodynamic interactions are suppressed. The only
tunable parameter is the friction $\gamma$, which is inversely
proportional to the single particle diffusion constant.

In order to introduce hydrodynamic interactions without simulating the
solvent explicitly, we used the lattice Boltzmann method~\cite{duenweg09a} on a
three-dimensional (3D) lattice with 19 velocity densities (D3Q19). In all
reported simulations with HIs, we used a
grid spacing of $g=1.0$.
The NVT ensemble was realized by the fluctuating LB algorithm~\cite{adhikari05a,
schiller08a}.  We treated the colloidal particles as
point particles that are coupled to the LB fluid via a friction
term with an adjustable friction constant $\gamma$~\cite{ahlrichs99a}. Note that
this friction constant is equivalent to
the friction in the Langevin dynamics, which is why we use the same
label $\gamma$. The no-slip boundaries modeling the fluid-wall
interaction were realized by the link bounce back
rule~\cite{ziegler93a}.  The strength of the hydrodynamic interactions
can be quantified by the hydrodynamic radius
\begin{equation}
 r_H=\frac{k_B T}{6 \pi \eta D^0},
  \label{equ2}
\end{equation}
where $r_H$ depends on the viscosity $\eta$ of the fluid as well as of
the single particle diffusion coefficient $D^0$, which is inversely
proportional to the friction constant $\gamma$. In contrast to other
methods for including HIs, such as DPD~\cite{hoogerbrugge92a, espanol95a},
LB methods allow the friction $\gamma$ to be tuned independently from
the viscosity $\eta$ of the fluid. The particle mobility is not simply the
inverse of $\gamma$, but also depends on the
viscosity $\eta$ and the lattice spacing of the LB grid due to
feedback from the moving fluid~\cite{ahlrichs99a, usta05a}. Also, the
back flow of the solvent introduces finite-size effects in a system with
periodic boundary conditions. 

The equations of motion of the Yukawa particles were integrated by a
Velocity-Verlet integrator~\cite{hinch75a}. If not otherwise stated, the
simulations were performed with $16,384$ particles in a box of size
$66\times 16\times 16$ confined by two planar walls located at $x=0.5$
and $x=65.5$. The time step of the Velocity Verlet integrator was
set to $dt=0.01$, the total simulation length $750,000$ time steps. The same
time step was also used for the LB fluid update, when applied. As basic length
we use the mean particle distance in the crystal phase $a=1.1$. The time
step is $d\tau^*=0.01\tau$, with $\tau=a\sqrt{k_B T/m_{p}}$, where $m_{p}$ is
the mass of the colloids. The viscosity is $\eta=0.8$ and the density of the
fluid is $\rho_{fl}=1.0$ as well as the particle density $\rho=1.0$. The
friction $\gamma$ varies between $0.5$ and $12.5$.

\begin{figure}
  \begin{center}
    \includegraphics[width=0.5\textwidth]{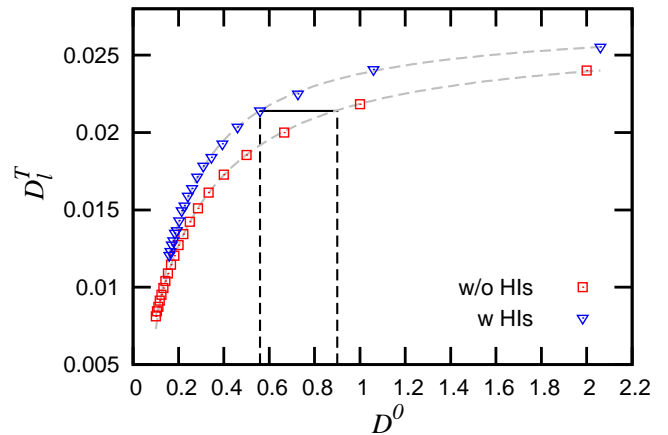}
    \caption{The diffusion coefficient of the tracer particles in the
      bulk $D^T_l$ as a function of the single particle diffusion
      coefficients $D^0$.
      (in simulation units). Red squares show results for the
      system without HIs and the blue triangles for the systems with
      HI. The gray dashed lines are a guide to eye. The black lines illustrate
      the matching of the diffusion coefficient. Two different single particle
      diffusion $D^0$ in the Langevin dynamics and the LB coupling result
      in the same long-time diffusion coefficient $D^T_l$.
    }
    \label{fig:3.1}
  \end{center}
\end{figure}

The effective diffusion constant depends not only on the applied
thermostat, but also on the interactions between the particles. In
order to set up comparable simulations, we therefore matched the
tracer diffusion coefficient in the bulk liquid. This was done by
measuring the MSD of tracer particles in pure bulk systems, both with and
without HIs. These measurements were done in 3D periodic systems consisting of
$16,384$ particles in a box of size $64\times 16\times 16$ without
confining walls. Figure~\ref{fig:3.1} shows the measured long-time
diffusion coefficient $D^T_l$ of tracer particles in the bulk calculated from
the MSD, as a function of the single particle diffusion coefficient $D^0$. The
simulations with and without HIs led to significantly different tracer diffusion
coefficients in the bulk, due to the fact that e.g. the single particle
diffusion in case of LB is not simply $D^0 \varpropto
1/\gamma$~\cite{ahlrichs99a}. Furthermore, there is no simple analytical rule to
calculate $D^T_l$ in case of a soft-sphere bulk systems. In order to match the
tracer particle diffusion coefficient $D^T_l$, we used different values of
$\gamma$ in the Langevin dynamics and the LB coupling. The black line in
Fig.~\ref{fig:3.1} illustrates this matching: in order to obtain the same
diffusion coefficient $D^T_l=0.015$, we have to apply $\gamma=4.0$ for the
Langevin dynamics and $\gamma=7.0$ for the LB coupling. In the following, we
always report data with matched tracer diffusion coefficients in the bulk, where
the given values of $\gamma$ are always the ones that apply to the LB coupling.

Using the matched tracer diffusion, we investigated the freezing of the
undercooled fluid confined between two planar walls. In order to distinguish the
liquid and the different solid phases, we used the Steinhardt order
parameter~\cite{steinhardt83a}. Investigations by Moroni \textit{et
al.}~\cite{moroni05a} showed that especially $q_4$ and $q_6$ are good
choices to determine whether cubic or hexagonal structures are present
in the system, respectively. In the following, we will focus on FCC
and BCC crystal structures, for which the $q_6$ order parameter is
well suited.  Due to the strong fluctuations in our system, we applied
an enhanced averaging method for the Steinhardt order parameter
$\overline{q}_6$, introduced by Lecher~\cite{lechner08a}. The
literature values are $\overline{q}_6(\mathrm{BCC})=0.408018$,
$\overline{q}_6(\mathrm{HCP})=0.42181$ and
$\overline{q}_6(\mathrm{LIQ})=0.161962$.

\begin{figure}[h]
  \begin{center}
    \includegraphics[width=0.5\textwidth]{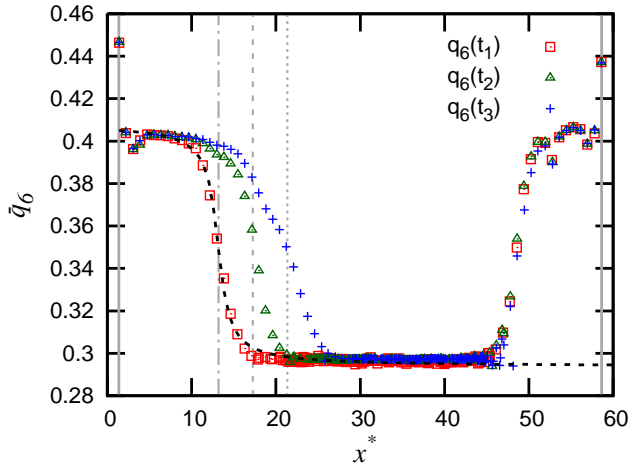}
    \caption{The figure shows our fitting procedure for s (position of
      the crystal front), from the function of $\overline{q}_6$ depending on
      $x^*=x/a$ for different times $t_1<t_2<t_3$. The gray dashed lines 
      indicate the computed front locations ($s_1<s_2<s_3$), while the black
      dashed line shows the fit for $s_1$ at time $t_1$. The symbols represent
the peaks      of the $\overline{q}_6$ order parameter.}
    \label{fig:3.2}
  \end{center}
\end{figure}

Figure~\ref{fig:3.2} shows the measured $\overline{q}_6$ of three 
snapshots taken at different times during a typical simulation run. Note that
the points only represent the peaks the distribution of
$\overline{q}_6$, since in the crystal, there is a strong layering
parallel to the wall. In between the peaks, the density drops nearly to
zero in the crystal, and consequently so does the order parameter. As
expected, the crystal starts with a HCP wall layer, followed by a BCC
crystal front that grows with time.  To evaluate the position of the
crystal front, we fitted the $\overline{q}_6$ peaks to a function of
shape $-h \cdot \arctan((x-s)/w)$, where $x$ is the $x$-position in
the simulation box, $h$ is the height
difference between $\overline{q}_6$ in the liquid and in the FCC phase, $w$ is
the width of the liquid-crystal transition region and $s$ is the position of
the crystal front. Note that $\overline{q}_6$ in the
liquid bulk is larger than the literature value, since we report only the peaks,
not the usual average value.

\begin{figure}
  \begin{center}
    \includegraphics[width=0.5\textwidth]{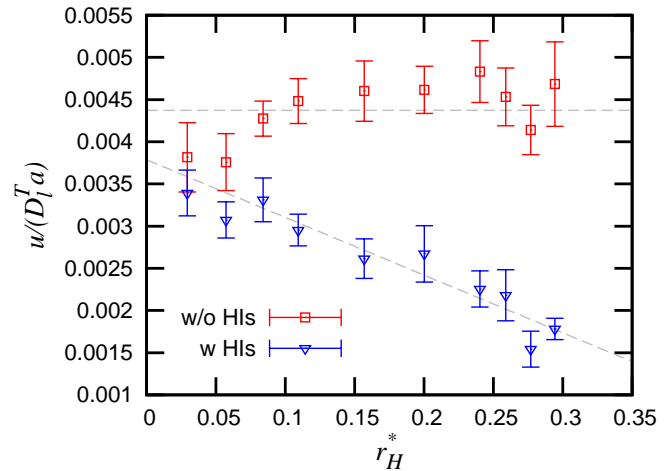}
    \caption{ Growth velocity $u$
      normalized by the bulk diffusion coefficient $D^T_l$ times the mean
      particle distance $a$ as a function of the relative
      hydrodynamic radius $r^*_H=r_H/a$. The red squares show the results
      for simulations without HIs, which are almost independent of $r^*_H$
      within the error bars. The blue triangles represent the results for
      simulations with HIs, which show a strong decay of the growth
      velocity with increasing hydrodynamic radius. The gray dashed lines are
      guidelines to eye. 
    }
    \label{fig:3.3}
  \end{center}
\end{figure}

After some initial time, the crystal grew very uniformly, so that we could
determine a constant growth velocity $u$ by a linear
fitting of the front position. Figure~\ref{fig:3.3} shows the measured
velocities $u$ as a function of the hydrodynamic radius $r_H$, which
we varied by changing the friction coefficient $\gamma$ and applying
the matching procedure described above. Every measurement represents
the mean growth velocity sampled from $24$ independent runs. For
hydrodynamic radii $r_H^* = r_H/a < 0.025$, where $a$ is the mean particle
distance in the crystal phase, the influence of HI is almost negligible as one
would expect. But already in case of moderate ratios $0.1 < r_H^* < 0.25$,
hydrodynamic interactions reduce the crystal growth velocity by up to a factor
of 3 at $r_H^*=0.25$. In case of no HIs the normalized growth velocity is
virtually constant, with a decay for small frictions due to improper
coupling to the thermostat.

\begin{figure}
  \begin{center}
    \includegraphics[width=0.5\textwidth]{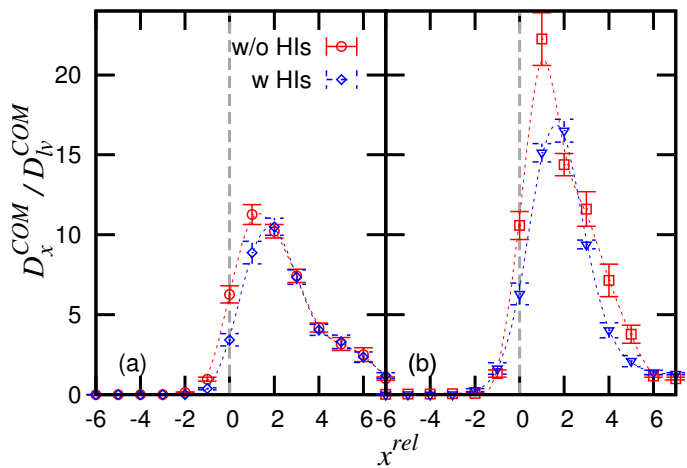}
    \caption{ Long-time diffusion coefficient of the center of mass in the
              direction of growth. Every value is the long-time
              diffusion coefficient of the center of mass for particles in a bin
              of size $b=2a$ normalized by the value of the Langevin simulation
              (w/o HIs). The front of the crystal is located at $x^{rel}=0$,
              while $x^{rel}=7$ is located in the pure bulk fluid phase. (a) The
              first two cases (I: red circles, II: blue rhombus) have a low
              $r^*_H=r_H/a=0.025$ ratio and show virtually similar behavior. (b)
              For moderate $r^*_H=0.25$, case IV (blue triangles) with HIs shows
              a different behavior from case III (red squares) without
              HIs, especially in the region in front of the crystal phase.
    }
    \label{fig:3.4}
  \end{center}
\end{figure}

In order to elucidate what causes this difference, we analyzed the
particle diffusion in the system relative to the actual position of the crystal
front. To accomplish this, we binned our system along the growth
direction into bins of width $b=2a$ and determined the long-time diffusion
coefficient $D^{COM}_x$ of the center of mass in the direction of growth, which
can be seen as a measure for the transport of particles towards the growing
crystal front. In Fig.~\ref{fig:3.4} $D^{COM}_x$ relative to the position of the
crystal front $x^{rel}$, normalized by the center of mass diffusion in the bulk
$D^{COM}_{lv}$ of the Langevin simulation is shown. The front of the crystal is
located at $x^{rel}=0$, while the pure bulk fluid phase is located at
$x^{rel}=7$ and the crystal phase is present for $x^{rel}<0$. As expected, the
long-time diffusion coefficients for the center of mass in the crystalline
region are almost zero, rise in the region of the crystal front, and settle off
to the liquid bulk value far away of the crystal front. The left-hand side of
Fig.~\ref{fig:3.4} shows the values for low $r^*_H=r_H/a=0.025$ ratio, which are
virtually the same for both systems: with and without HIs. However, on the
right-hand side $r^*_H=0.25$ is shown, corresponding to moderate hydrodynamic
coupling, for which a significantly reduction of the diffusion is present in
case of HIs. Especially, nearby the crystal front a drop down of more than a
factor of two in the diffusion coefficient occurs, corresponding to a
drastically reduced transport towards the crystal.
We believe that the reduced diffusion towards the crystal front in
the LB system is caused by
hydrodynamic screening due to the presence of the wall-like crystal surface.
Hydrodynamic screening nearby walls and their influence on the diffusion
coefficients parallel and perpendicular to the wall has been reported by Feitosa
\emph{et al.}~\cite{feitosa91a} and von Hansen \emph{et
al.}~\cite{vonhansen11a}.

Our simulations show that hydrodynamic interactions have a strong
influence on crystallization, even at moderate hydrodynamic radii. Similar
finding has been reported by Schilling \emph{et al.}~\cite{radu13a} as
well. The effects arise mainly on the particle transport towards the crystal
front, which are in particular important for nucleation processes. This puts
the common assumption into doubt that hydrodynamic
interactions can be ignored when studying crystallization or
nucleation in suspensions. At least in Yukawa suspensions, these processes do
not seem to be quasi-static, and the often drawn analogy to true melts
might not be true.

We thank the staff at the Institute for Computational Physics for useful
discussions and acknowledge financial support from the German Science Foundation
(DFG) through SFB716 and the cluster of excellence SimTech.

\end{document}